\documentclass[english,aps,prb,amsmath,twocolumn, amssymb,superscriptaddress,floatfix]{revtex4-1}
\usepackage[T1]{fontenc}
\usepackage[latin9]{inputenc}
\usepackage{color}
\usepackage{amsmath}
\usepackage{amssymb}
\usepackage{graphicx}

\makeatletter
\providecommand{\tabularnewline}{\\}
\providecommand{\tabularnewline}{\\}
\usepackage{latexsym}
\usepackage{graphics}
\usepackage{epsfig}
\usepackage{amsmath}
\usepackage{amsopn}
\usepackage{amsfonts}
\usepackage{wasysym}

\usepackage{nicefrac}
\expandafter\let\csname equation*\endcsname\relax
\expandafter\let\csname endequation*\endcsname\relax

\renewcommand{\Re}{\mathfrak{Re}}
\renewcommand{\Im}{\mathfrak{Im}}

\makeatother

\def\bibsection{\section*{References}}

\usepackage{babel}
\begin{document}

\title{Reversing the direction of heat flow using quantum correlations}

\author{Kaonan Micadei}

\thanks{These authors contributed equally to this work.}

\affiliation{Centro de Ci\^encias Naturais e Humanas, Universidade Federal do ABC,
Avenida dos Estados 5001, 09210-580 Santo Andr\'e, S\~ao Paulo, Brazil}
\affiliation{Institute for Theoretical Physics I, University of Stuttgart, D-70550 Stuttgart, Germany}

\author{John P. S. Peterson}

\thanks{These authors contributed equally to this work.}

\affiliation{Centro Brasileiro de Pesquisas F\'isicas, Rua Dr. Xavier Sigaud 150,
22290-180 Rio de Janeiro, Rio de Janeiro, Brazil}

\author{Alexandre M. Souza}

\affiliation{Centro Brasileiro de Pesquisas F\'isicas, Rua Dr. Xavier Sigaud 150,
22290-180 Rio de Janeiro, Rio de Janeiro, Brazil}

\author{Roberto S. Sarthour}

\affiliation{Centro Brasileiro de Pesquisas F\'isicas, Rua Dr. Xavier Sigaud 150,
22290-180 Rio de Janeiro, Rio de Janeiro, Brazil}

\author{Ivan S. Oliveira}

\affiliation{Centro Brasileiro de Pesquisas F\'isicas, Rua Dr. Xavier Sigaud 150,
22290-180 Rio de Janeiro, Rio de Janeiro, Brazil}

\author{Gabriel T. Landi}

\affiliation{Instituto de F\'isica, Universidade de S\~ao Paulo, C.P. 66318, 05315-970
S\~ao Paulo, SP, Brazil}

\author{Tiago B. Batalh\~ao}

\address{Singapore University of Technology and Design, 8 Somapah Road, Singapore
487372}

\address{Centre for Quantum Technologies, National University of Singapore,
3 Science Drive 2, Singapore 117543}

\author{Roberto M. Serra}

\affiliation{Centro de Ci\^encias Naturais e Humanas, Universidade Federal do ABC,
Avenida dos Estados 5001, 09210-580 Santo Andr\'e, S\~ao Paulo, Brazil}

\affiliation{Department of Physics, University of York, York YO10 5DD, United
Kingdom}

\author{Eric Lutz}
%

\affiliation{Institute for Theoretical Physics I, University of Stuttgart, D-70550 Stuttgart, Germany}

\begin{abstract}
Heat spontaneously flows from hot to cold in standard thermodynamics. However, the latter theory presupposes the absence of initial correlations between interacting systems. We here experimentally demonstrate the reversal of heat flow for two quantum correlated spins-1/2, initially prepared in local thermal states at different effective temperatures, employing a Nuclear Magnetic Resonance setup. We observe a spontaneous energy flow from the cold to the hot system. This process is enabled by a trade off between correlations and entropy that we quantify with information- theoretical quantities. These results highlight the subtle interplay of quantum mechanics, thermodynamics and information theory. They further provide a mechanism to control heat on the microscale.
\end{abstract}
\maketitle

According to Clausius, heat spontaneously flows from a hot body to a cold body \cite{cla79}.  At a phenomenological level, the second law  of thermodynamics  associates such irreversible behavior with a nonnegative mean entropy production \cite{cal85}. On the other hand, Boltzmann related it to specific initial conditions of the microscopic dynamics  \cite{bol96,leb93,zeh07}. Quantitative experimental confirmation of this conjecture
has recently been obtained for a driven classical Brownian particle
and for an electrical RC circuit \cite{and07}, as well as for a driven
quantum spin \cite{bat15}, and a driven quantum dot \cite{hof17}.
These experiments have been accompanied by a surge of theoretical
studies on  classical and quantum irreversibility \cite{par08,jev12,jen10,ber17,fen08,mac09,par09,jar11,rol15,ved16,dre17}.
It has in particular been shown that a preferred direction of average
behavior may be discerned irrespective of the size of the system \cite{cam11}.

Initial conditions not only induce irreversible heat flow, they also determine the direction of the heat current. The observation of the average positivity of
the entropy production in nature is often explained by the low entropy
value of the initial state \cite{bol96}. This opens the possibility
to control or even reverse the direction of heat flow depending on the initial
conditions. In standard thermodynamics, systems are assumed to be
uncorrelated before thermal contact. As a result, according to the
second law heat will flow from the hot object to the cold object. However,
it has been theoretically suggested that for quantum correlated local
thermal states, heat might flow from the cold to the hot system, thus
effectively reversing its direction \cite{par08,jev12,jen10,ber17}. This phenomenon has been predicted to occur in general multidimensional bipartite systems \cite{par08,jev12}, including the limiting case of two simple qubits \cite{jev12}, as well as in multipartite systems \cite{jen10}.

Here we report the experimental demonstration of the reversal of heat flow for two initially quantum correlated qubits (two spin-1/2
systems) prepared in local thermal states at different effective temperatures
employing Nuclear Magnetic Resonance (NMR) techniques \cite{oli07,van04}.
Allowing thermal contact between the  qubits, we track the evolution
of the global state with the help of quantum state tomography 
\cite{oli07}. We experimentally determine the energy change of each
spin and the variation of their mutual information \cite{nie00}.
For initially correlated systems, we observe a spontaneous heat current
from the cold to the hot spin and show that this process is made possible
by a decrease of their mutual information. The second
law for the isolated two-spin system is therefore verified. However,
the standard second law in its local form apparently fails to apply
to this situation with initial quantum correlations. We further establish
the nonclassicallity of the initial correlation by evaluating its
non-zero geometric quantum discord, a measure of quantumness \cite{dak10,gir12}.
We finally theoretically derive and experimentally investigate an
expression for the heat current that reveals the trade off between
information and entropy.

\begin{figure*}[ht]
\centering \includegraphics[width=0.95\textwidth]{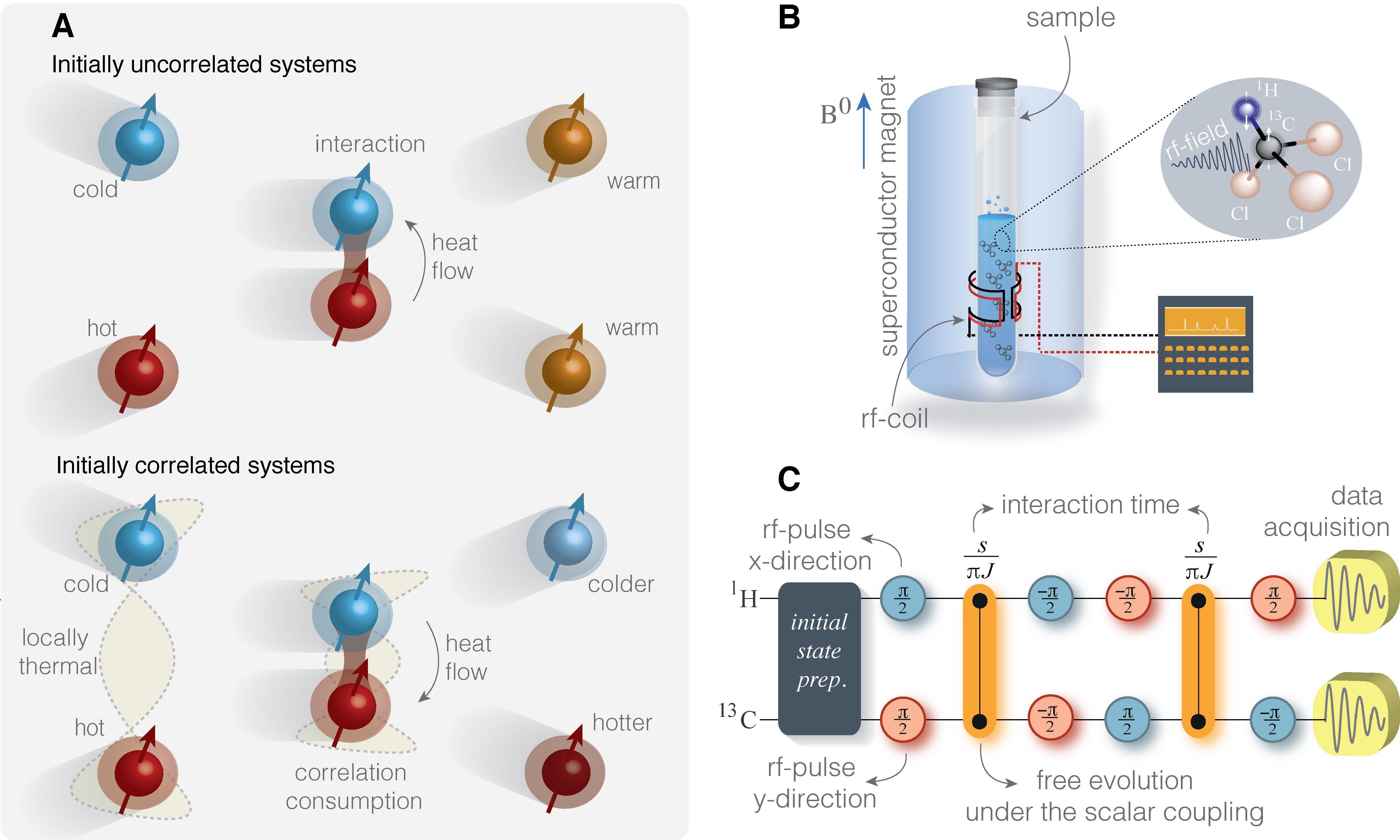} \caption{\textbf{Schematic of the experimental setup}. \textbf{A} Heat flows
from the hot to the cold spin (at thermal contact) when both are initially
uncorrelated. This corresponds to  standard thermodynamic. For initially quantum correlated spins, heat is spontaneously
transferred from the cold to the hot spin. The direction of heat flow is here
reversed. \textbf{B} View of the magnetometer used in our NMR experiment.
A superconducting magnet, producing a high intensity magnetic field
$(B_{0})$ in the longitudinal direction, is immersed in a thermally
shielded vessel in liquid He, surrounded by liquid N in another vacuum
separated chamber. The sample is placed at the center of the magnet
within the radio frequency coil of the probe head inside a $5$mm
glass tube. \textbf{C} Experimental pulse sequence for the partial
thermalization process. The blue (black) circle represents $x$ ($y$)
rotations by the indicated angle. The orange connections represents
a free evolution under the scalar coupling, $\mathcal{H}_\text{J}^{\text{HC}}=({\pi\hbar}/{2})J\sigma_{z}^{\text{H}}\sigma_{z}^{\text{C}}$,
between the $^{1}$H and $^{13}$C nuclear spins during the time indicated
above the symbol. We have performed 22 samplings of the interaction
time $\tau$ in the interval 0 to $2.32$~ms.}
\label{fig1} 
\end{figure*}

\vspace{.1in}
\noindent {\bf \large{\textsf{Results}}}\\
\textbf{Experimental system.} NMR offers an exceptional degree of preparation, control, and measurement
of coupled nuclear spin systems \cite{oli07,van04}. It has for this
reason become a premier tool for the study of quantum thermodynamics
\cite{bat15,bat14,cam16}. In our investigation, we consider two nuclear
spins-1/2, in the $^{13}$C and $^{1}$H nuclei of a $^{13}$C-labeled
CHCl$_{3}$ liquid sample diluted in Acetone-d6 (Fig.~\ref{fig1} B).
The sample is placed inside a superconducting magnet that produces
a longitudinal static magnetic field (along the positive $z$-axis)
and the system is manipulated by time-modulated transverse radio-frequency
(rf) fields. We study processes in a time interval of few milliseconds
which is much shorter than any relevant decoherence time of the system
(of the order of few seconds) \cite{bat14}. The dynamics of the combined
spins in the sample is thus effectively closed and the total energy
is conserved to an excellent approximation. Our aim is to study the
heat exchange between the $^{1}$H (system A) and $^{13}$C (system
B) nuclear spins under a partial thermalization process in the
presence of initial correlations (Fig.~\ref{fig1} A). Employing a sequence of transversal
rf-field and longitudinal field-gradient pulses, we prepare an initial
state of both nuclear spins (A and B) of the form, 
\begin{equation}
\rho_\text{AB}^{0}=\rho_\text{A}^{0}\otimes\rho_\text{B}^{0}+\chi_\text{AB},\label{eq:1}
\end{equation}
where $\chi_\text{AB}=\alpha|01\rangle\!\langle10|+\alpha^{*}|10\rangle\!\langle01|$
is a correlation term and $\rho_{i}^{0}=\exp(-\beta_{i}\mathcal{H}_{i})/\mathcal{Z}_{i}$
a thermal state at inverse temperature $\beta_{i}=1/(k_{B}T_{i})$,
$i=(\text{A,B})$, with $k_{B}$ the Boltzmann constant. The state $\left|0\right\rangle $
($\left|1\right\rangle $) represents the ground (excited) eigenstate
of the Hamiltonian $\mathcal{H}_{i}$, and $\mathcal{Z}_{i}=\mathrm{Tr}_{i}\,\exp(-\beta_{i}\mathcal{H}_{i})$
is the partition function. The individual nuclear spin Hamiltonian,
in a double-rotating frame with the nuclear spins ($^{1}\text{H}$
and $^{13}\text{C}$) Larmor frequency, may be written
as $\mathcal{H}_{i}=h\nu_{0}\left(\mathbf{1}-\sigma_{z}^{i}\right)/2$,
with $\nu_{0}=1$~kHz effectively determined by a nuclei rf-field
offset. In Eq.~(\ref{eq:1}), the coupling strength should satisfy
$|\alpha|\leq\exp[-h\nu_{0}(\beta_\text{A}+\beta_\text{B})/2]/(\mathcal{Z}_\text{A}\mathcal{Z}_\text{B})$
to ensure positivity. We consider two distinct cases: for $\alpha=0$
the  spins are initially uncorrelated as assumed in standard thermodynamics,
while for $\alpha\neq0$ the joint state is initially correlated.
We note that since $\mathrm{Tr}_{i}\,\chi_\text{AB}=0$, the two spins 
are locally always in a thermal Gibbs state in both situations. As a result, thermodynamic quantities, such as temperature, internal energy, heat and entropy, are well defined. A partial thermalization between
the qubits is described by the effective (Dzyaloshinskii-Moriya) interaction
Hamiltonian, $\mathcal{H}_\text{AB}^\text{eff}=({\pi\hbar}/{2})J(\sigma_{x}^\text{A}\sigma_{y}^{B}-\sigma_{y}^\text{A}\sigma_{x}^\text{B})$,
with $J=215.1$~Hz \cite{sca02,zim02}, which can be easily realized experimentally. We implement the corresponding
evolution operator, $\mathcal{U}_{\tau}=\exp(-i\tau\mathcal{H}_\text{AB}^\text{eff}/\hbar)$,
by combining free evolutions under the natural hydrogen-carbon scalar
coupling and rf-field rotations (Fig.~\ref{fig1} C). We further stress that the  correlation term  should satisfy  $[\chi_\text{AB},\mathcal{H}_\text{AB}^\text{eff}]\neq0$ for  the heat flow reversal to occur (Supplementary Information).

\begin{figure*}[t]
\centering \includegraphics[width=1.9\columnwidth]{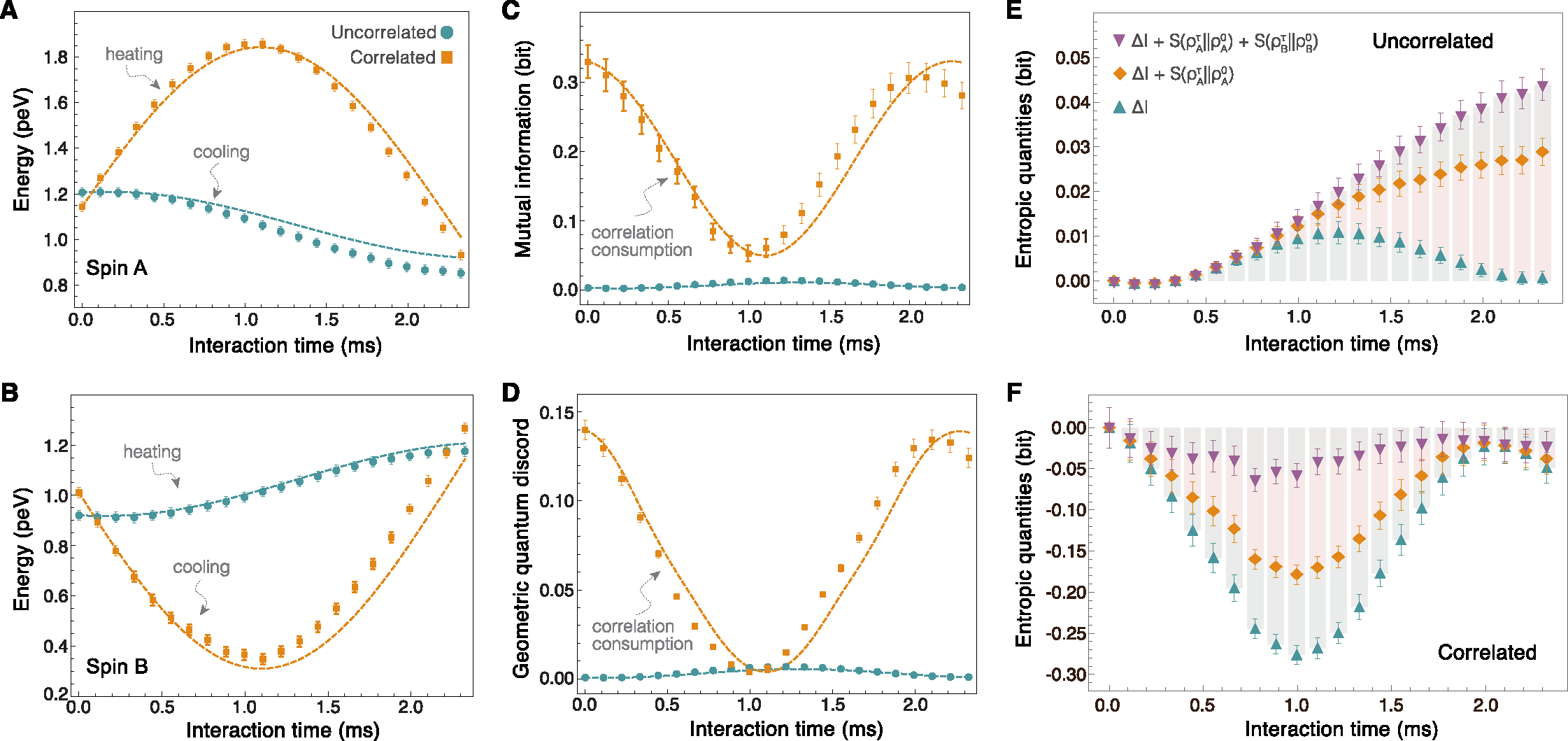} \caption{\textbf{Dynamics of heat, correlations, and entropic quantities.}
\textbf{A} Internal energy of qubit A along the partial thermalization
process. \textbf{B} Internal energy of qubit B. In the absence
of initial correlations, the hot qubit A cools down and the cold
qubit B heats up (cyan circles in panel \textbf{A} and \textbf{B}).
By contrast, in the presence of initial quantum correlations, the
heat current is reversed as the hot qubit $A$ gains and the cold
qubit B loses energy (orange squares in panel \textbf{A} and \textbf{B}).
This reversal is made possible by a decrease of the mutual information
\textbf{C} and of the geometric quantum discord \textbf{D}. Different
entropic contributions to the heat current (5) in the uncorrelated
\textbf{E} and uncorrelated \textbf{F} case. Reversal occurs when
the negative variation of the mutual information, $\Delta I(A{:}B)$,
compensates the positive entropy productions, $S(\rho_\text{A}^{\tau}||\rho_\text{A})$
and $S(\rho_\text{B}^{\tau}||\rho_\text{B})$, of the respective qubits. The
symbols represent experimental data and the dashed lines are numerical
simulations. Error bars were estimated by a Monte Carlo sampling from the standard 
deviation of the measured data (Supplementary Information).}
\label{fig2} 
\end{figure*}

 \noindent \textbf{Thermodynamics.} In macroscopic thermodynamics, heat is  defined as the energy exchanged between to large bodies at different temperatures \cite{cal85}. This notion has been successfully extended to small systems initially prepared in thermal Gibbs states \cite{jar04}, including qubits \cite{jev12}. Since the interaction  Hamiltonian commutes with the total Hamiltonian of the two qubits, $[\mathcal{H}_\text{A}+H_\text{B}, \mathcal{H}_\text{AB}^\text{eff}] = 0$, the thermalization operation does not perform  any work on the spins \cite{bar15}. As a result, the mean energy is constant in time and the heat absorbed by one qubit is given
by its internal energy variation along the dynamics, $Q_{i}=\Delta E_{i}$,
where $E_{i}=\mathrm{Tr}_{i}\,\rho_{i}\mathcal{H}_{i}$ is the $z$-component
of the nuclear spin magnetization. For the combined system, the two
heat contributions satisfy \cite{par08,jen10,jev12},
\begin{equation}
\beta_\text{A}Q_\text{A}+\beta_\text{B}Q_\text{B}\geq\Delta I(A{:}B),
\end{equation}
where $\Delta I(A{:}B)$ is the change of mutual information between
$A$ and $B$. The mutual information, defined as $I(A{:}B)=S_\text{A}+S_\text{B}-S_\text{AB}\geq0$,
is a measure of the total correlations between two systems \cite{nie00},
where $S_{i}=-\mathrm{Tr}_{i}\rho_{i}\ln\rho_{i}$ is the von Neumann
entropy of state $\rho_{i}$. Equation (2) follows from the unitarity of the global dynamics and the Gibbs form of the initial spin states. For initially uncorrelated spins, the
initial mutual information is zero. As a result, it can only increase
during thermalization, $\Delta I(A{:}B)\geq0$. Noting that $Q_\text{A}+Q_\text{B}=0$
for the isolated bipartite system, we find \cite{par08,jen10,jev12},
\begin{equation}
Q_\text{B}(\beta_\text{B}-\beta_\text{A})\geq0\quad\text{(uncorrelated)}.
\end{equation}
Heat hence flows from the hot to the cold spin, $Q_{\text{B}}>0$ if $T_{\text{A}}\geq T_{\text{B}}$.
This is the standard second law. By contrast, for initially correlated
qubits, the mutual information may decrease during the thermal contact
between the spins. In that situation, we may have \cite{par08,jen10,jev12},
\begin{equation}
Q_\text{B}(\beta_\text{B}-\beta_\text{A})\leq0\quad\text{(correlated)}.
\end{equation}
Heat flows in this case from the cold to the hot qubit: the energy current
is reversed. We quantitatively characterize the occurrence
of such reversal by computing the heat flow  at any time $\tau$,
obtaining (Methods), 
\begin{equation}
\Delta\beta Q_\text{B}=\Delta I(A{:}B)+S(\rho_\text{A}^{\tau}||\rho^0_\text{A})+S(\rho_\text{B}^{\tau}||\rho^0_\text{B}),\label{eq:5}
\end{equation}
where $\Delta\beta=\beta_\text{B}-\beta_{A}\geq0$ and $S(\rho_{i}^{\tau}||\rho_{i})=\mathrm{Tr}_{i}\,\rho_{i}^{\tau}(\ln\rho_{i}^{\tau}-\ln\rho_{i})\geq0$
denotes the relative entropy \cite{nie00} between the evolved $\rho_\text{A(B)}^{\tau}=\mathrm{Tr}_\text{B(A)}\mathcal{U}_{\tau}\rho_\text{AB}^{0}\mathcal{U}_{\tau}^{\dagger}$
and the initial $\rho_\text{A(B)}^{0}$ reduced states. The latter quantifies
the entropic distance between the state at time $\tau$ and the initial
thermal state. {It can be interpreted as the entropy production associated with the irreversible heat transfer, or  to the entropy produced during the ensuing relaxation to the initial thermal state~\cite{def10,def11}}. According to Eq.~(\ref{eq:5}), the
direction of the {energy current} is therefore reversed whenever the decrease
of mutual information compensates the entropy production. The fact
that initial correlations may be used to decrease entropy has first
been emphasized by Lloyd \cite{llo89} and further investigated in
Refs.~\cite{sag12,kos14}. {Heat flow reversal has recently been interpreted as a refrigeration process driven by the work potential stored in the correlations \cite{ber17}. In that context, Eq.~\eqref{eq:5} can be seen as a generalized Clausius inequality due to the positivity of the relative entropies. The coefficient of performance of such a refrigeration is then at most that of Carnot \cite{ber17}.}

In our experiment, we prepare the two-qubit system in an initial state
of the form (1) with effective spin temperatures $\beta_\text{A}^{-1}=4.66\pm0.13$~peV
($\beta_\text{A}^{-1}=4.30\pm0.11$~peV) and $\beta_\text{B}^{-1}=3.31\pm0.08$~peV
($\beta_\text{B}^{-1}=3.66\pm0.09$~peV) for the uncorrelated (correlated)
case $\alpha=0.00\pm0.01$ ($\alpha=-0.19\pm0.01$) (Supplementary Information). {The value of $\alpha$ was chosen to  maximize the current reversal}. In
order to quantify the quantumness of the initial correlation in the
{correlated} case, we consider the normalized geometric discord, defined
as $D_\text{g}=\min_{\psi\in\mathcal{C}}2\|\rho-\psi\|^{2}$ where $\mathcal{C}$
is the set of all states classically correlated \cite{dak10,gir12}.
The geometric discord {has a simple closed form expression} for two qubits that can be directly evaluated from
the measured QST data (Supplementary Information). We find the nonzero value $D_\text{g}=0.14\pm0.01$
for the initially correlated state prepared in the experiment. 

We experimentally reconstruct the global two-qubit density operator using quantum state tomography \cite{oli07} and evaluate the changes of  internal energies of each qubit, of mutual
information, and of geometric quantum discord during thermal contact
(Figs.~\ref{fig2} A to F). We observe the standard second law in the absence of initial correlations ($\alpha\simeq0$), i.e., the
hot qubit $A$ cools down, $Q_\text{A}<0$, while the cold qubit $B$ heats
up, $Q_\text{B}>0$ (circles symbols in Figs.~\ref{fig2} A and B). At
the same time, the mutual information and the geometric quantum discord
increase, as correlations build up following the thermal interaction
(circles symbols in Figs.~\ref{fig2} C and D). The situation changes
dramatically in the presence of initial quantum correlations ($\alpha\neq0$):
the {energy current} is here reversed in the time interval, $0<\tau<2.1$~ms,
as heat flows from the cold to the hot spin, $Q_\text{A}=-Q_\text{B}>0$ (squares
symbols in Figs.~\ref{fig2} A and B). This reversal is accompanied
by a decrease of mutual information and geometric quantum discord
(squares symbols in Fig.~\ref{fig2} C and D). In this case, quantum
correlations are converted into energy and used to switch the direction
of the heat flow, in an apparent violation of the second law. Correlations
reach their minimum at around $\tau\approx1.05$~ms, after which
they build up again. Once they have passed their initial value at
$\tau\approx2.1$~ms, energy is transferred in the expected direction,
from hot to cold. In all cases, we obtain good agreement between experimental
data (symbols) and theoretical simulations (dashed lines). Small discrepancies
seen as time increases are mainly due to inhomogeneities in the control
fields.

The experimental investigation of Eq.~(\ref{eq:5}) as a function
of the thermalization time is presented in Fig.~\ref{fig2} E and
F. While the relative entropies steadily grow in the absence of initial
correlations, they exhibit an increase up to $1.05$~ms followed
by a decrease in presence of initial correlations. The latter behavior
reflects the pattern of the qubits already seen in Fig.~\ref{fig2}
A and B, for the average energies. We note in addition a positive
variation of the mutual information in the uncorrelated case and a
large negative variation in the correlated case. The latter offsets
the increase of the relative entropies and enables the reversal of
the heat current. These findings provide direct experimental evidence
for the trading of quantum mutual information and entropy production.

\vspace{.1in}
\noindent {\bf \large{\textsf{Discussion}}}\\
We have observed the reversal of the energy flow between two quantum-correlated qubits with different effective temperatures, associated with the respective populations of the two levels. Such effect has been predicted to exist in  general multidimensional  systems \cite{par08,jev12,jen10}.
By revealing the fundamental influence of initial quantum correlations
on {the direction of thermodynamic processes, which  Eddington
has called the arrow of time \cite{edd28}}, our experiment highlights the subtle interplay of
quantum mechanics, thermodynamics and information theory.  Initial conditions
thus not only break the time-reversal symmetry of the otherwise reversible dynamics, they also determine the direction of a process.
Our findings further
emphasize the limitations of the standard local formulation of the
second law for initially correlated systems and offers at the same
time a novel mechanism to control heat on the microscale. They additionally
establish that the arrow of time is not an absolute but a relative
concept that depends on the choice of initial conditions. {We
have observed the reversal of the energy current for the case of two spins which never fully thermalize due to their finite size. However, their dynamics is identical to that of a thermalization map during the duration of the experiment (Methods), a process  we have labeled partial thermalization for this reason. Furthermore,  numerical
simulations show that reversals may also occur for a spin interacting
with larger spin environments which induce thermalization \cite{sup}}. Hence, an anomalous heat
current does not seem to be restricted to  microscopic systems.
The precise scaling of this effect with the system size is an interesting
subject for future experimental and theoretical investigations. Our
results on the reversal of the thermodynamic arrow of time might also have stimulating
consequences on the cosmological arrow of time \cite{llo89}.

\vspace{.25in}

\noindent {\bf \large{\textsf{Methods}}}\\
{\small
\textbf{Experimental setup.} The liquid sample consist of $50$~mg
of $99$\% $^{13}\text{C}$-labeled $\text{CHCl}_{3}$ (Chloroform)
diluted in $0.7$~ml of $99.9$\% deutered Acetone-d6, in a flame
sealed Wildmad LabGlass 5 mm tube. Experiments were carried out in
a Varian $500$~MHz Spectrometer employing a double-resonance probe-head
equipped with a magnetic field gradient coil. The sample is very diluted
such that the intermolecular interaction can be neglected, in this
way the sample can be regarded as a set of identically prepared pairs
of spin-1/2 systems. The superconducting magnet (illustrated in Fig.~1B
of the main text) inside of the magnetometer produces a strong intensity
longitudinal static magnetic field (whose direction is taken to be
along the positive $z$ axes), $B_{0}\approx11.75$~T. Under this
filed the $^{1}\text{H}$ and $^{13}\text{C}$ Larmor frequencies
are about $500$~MHz and $125$~MHz, respectively. The state of
the nuclear spins are controlled by time-modulated rf-field pulses
in the transverse ($x$ and $y$) direction and longitudinal field
gradients.

Spin-lattice relaxation times, measured by the inversion recovery
pulse sequence, are $(\mathcal{T}_{1}^\text{H},\mathcal{T}_{1}^\text{C})=(7.42,11.31)$~s.
Transverse relaxations, obtained by the Carr-Purcell-Meiboom-Gill
(CPMG) pulse sequence, have characteristic times $(\mathcal{T}_{2}^{*\text{H}},\mathcal{T}_{2}^{*\text{C}})=(1.11,\:0.30)$~s.
The total experimental running time, to implement the partial spin
thermalization, is about $2.32$\textbf{~}ms, which is considerably
smaller than the spin-lattice relaxation and therefore decoherence
can be disregarded.

\noindent\textbf{Heat current between initially correlated systems}.
We here derive the  expression (5) for the heat flow between initially correlated systems
A and B. For an initial thermal state $\rho_{i}^{0}=\exp(-\beta_{i}\mathcal{H}_{i})/\mathcal{Z}_{i}$,
$(i=A,B)$, the relative entropy between the evolved $\rho_\text{A(B)}^{\tau}=\mathrm{Tr}_\text{B(A)}\mathcal{U}_{\tau}\rho_\text{AB}^{0}\mathcal{U}_{\tau}^{\dagger}$
and the initial $\rho_{i}^{0}$ marginal states reads, 
\begin{equation}
S(\rho_{i}^{\tau}\|\rho_{i}^{0})=-S(\rho_{i}^{\tau})+\beta_{i}\,\mathrm{Tr}_{i}\,(\rho_{i}^{\tau}\mathcal{H}_{i})+\ln\mathcal{Z}_{i},
\end{equation}
where $S(\rho_{i}^{\tau})$ is the von Neumann entropy of the state
$\rho_{i}^{\tau}$. Noting that as $S(\rho_{i}^{0}||\rho_{i}^{0})=0$,
one can write, 
\begin{equation}
S(\rho_{i}^{\tau}\|\rho_{i}^{0})=S(\rho_{i}^{\tau}\|\rho_{i}^{0})-S(\rho_{i}^{0}||\rho_{i}^{0})=-\Delta S_{i}+\beta_{i}\Delta E_{i},
\end{equation}
with the variation in the von Neumann entropy, given by $\Delta S_{i}=S(\rho_{i}^{\tau})-S(\rho_{i}^{0})$
and the internal energy change of the $i$-th subsystem defined as
$\Delta E_{i}=\mathrm{Tr}_{i}\,\rho_{i}^{\tau}\mathcal{H}_{i}-\mathrm{Tr}_{i}\,\rho_{i}^{0}\mathcal{H}_{i}$.
Energy conservation for the combined isolated system (AB) further
implies that $\Delta E_\text{A}=-\Delta E_\text{B}=Q_\text{B}$, for constant interaction
energy. As a result, we obtain, 
\begin{equation}
Q_\text{B}\,\Delta\beta=\Delta I(A{:}B)+S\big(\rho_\text{A}^{\tau}\|\rho_\text{A}^{0}\big)+S\big(\rho_\text{B}^{\tau}\|\rho_\text{B}^{0}\big),
\end{equation}
where $\Delta\beta=\beta_\text{A}-\beta_\text{B}$ is the inverse temperature
difference and $\Delta S_\text{A}+\Delta S_\text{B}=\Delta I(A{:}B)$ holds
since the combined system (AB) is isolated.

\noindent {\textbf{Partial thermalization.} We further show that the spin dynamics  is given by a thermalization map during the duration of the experiment. From the local point of view of each individual nuclear spin (when the spins are initially uncorrelated), the evolution operator $\mathcal{U}_{\tau}=\exp(-i\tau\mathcal{H}_\text{AB}^\text{eff}/\hbar)$, with $\tau \in [0,(2 J)^{-1}]$, has the effect of a linear non-unitary map  $\mathcal{E}(\rho_i)=\mathrm{Tr}_{k}\left( \mathcal{U}_{\tau} \rho_\text{A}^0\otimes \rho_\text{B}^0 \mathcal{U}_{\tau} ^{\dagger}\right)$ on the spin $i$, which can be represented as, 
\begin{equation}
    \label{eq:kraus}
    \mathcal{E}(\rho_i) = \sum_{j=1}^4 K_j \rho_i^0 K_j^\dagger
\end{equation}
where $i=A, k=B$ or $i=B, k=A$. The Kraus corresponding operators $K_j$, with $j=(1,...,4)$, are given by,
\begin{eqnarray}
        K_1 &=& \sqrt{1-p}
        \begin{bmatrix}
            1 & 0 \\
            0 & \cos(\pi J \tau)
        \end{bmatrix} ,\\
        K_2 &=& \sqrt{1-p}
        \begin{bmatrix}
            0 & \sin(\pi J \tau) \\
            0 & 0
        \end{bmatrix} ,\\
         K_3 &=& \sqrt{p}
        \begin{bmatrix}
            \cos(\pi J \tau) & 0 \\
            0 & 1
        \end{bmatrix} ,\\
        K_4 &=& \sqrt{p}
        \begin{bmatrix}
            0 & 0 \\
            -\sin(\pi J \tau) & 0
        \end{bmatrix} ,
\end{eqnarray}
where $p$ is the population of the excited state in the other spin at time $\tau$.
In the time window of the experiment, $\pi J \tau$ varies between zero and $\pi/2$. In this way the transformation \eqref{eq:kraus} is equivalent to the generalized amplitude damping \cite{Srikanth2008} which is the Kraus map for the thermalization of a single spin-$1/2$ system. Therefore, from the local point of view and in the absence of initial correlations, the interaction implemented in the experiment is indistinguishable from a thermalization map for $\tau \in [0,(2 J)^{-1}]$.}}

\vspace{.3 in}
\noindent{\bf \large{\textsf{Acknowledgements}}}\\
 We acknowledge financial support from UFABC, CNPq, CAPES, FAPERJ, and FAPESP. R.M.S. gratefully acknowledges financial support from the Royal Society through the Newton Advanced Fellowship scheme (Grant no. NA140436) and the technical support from the Multiuser Experimental Facilities of UFABC. A.M.S. acknowledges support from the Brazilian agency FAPERJ (203.166/2017). K. M. acknowledges CAPES and DAAD for financial support. E.L. acknowledges support from the German Science Foundation (DFG) (Grant no. FOR 2724). This research was performed as part of the Brazilian National Institute of Science and Technology for Quantum Information (INCT-IQ).
\\
\\
\hspace{-.1 in}
\noindent{\bf \large{\textsf{Author contributions}}}\\
K.M., J.P.S.P., T.B.B., R.S.S., and R.M.S. designed the experiment, J.P.S.P., A.M.S., R.S.S., I.S.O., and R.M.S. performed the experiment, K.M., G.T.L., and E.L. contributed to the theory. All authors contributed to analyzing the data and writing the paper.
\\
\\
\hspace{-.1 in}
\noindent{\bf \large{\textsf{Data availability}}}\\
The datasets generated during and/or analysed during the current study are available from serra@ufabc.edu.br on reasonable request.
\\
\\
\hspace{-.1 in}
\noindent{\bf \large{\textsf{Competing interests}}}\\
The authors declare no competing interests..
\\


\section*{Supplementary Material}

\global\long\def\beginsupplement{%
\setcounter{table}{0} \global\long\global\long\global\long\def\thetable{S\Roman{table}}
\setcounter{figure}{0} \global\long\global\long\global\long\def\thefigure{S\arabic{figure}}
\setcounter{equation}{0} \global\long\global\long\global\long\def\theequation{S\arabic{equation}}
}
 \beginsupplement
 
\textbf{Supplementary Note 1: Initial State Preparation} 
\vspace{0.1 in}

The initial state of the nuclear spins is prepared by spatial average
techniques~\cite{bat15,oli07,bat14}, being the $^{1}\text{H}$ and
$^{13}\text{C}$ nuclei prepared in local pseudo-thermal states with
the populations (in the energy basis of $\mathcal{H}_{0}^{\text{H}}$
and $\mathcal{H}_{0}^{\text{C}}$) and corresponding local spin temperatures
displayed in Supplementary Table~SI. The initial correlated state is prepared through
the pulse sequence depicted in Supplementary Figure~S1.

\begin{figure}[ht]
\includegraphics[scale=0.38]{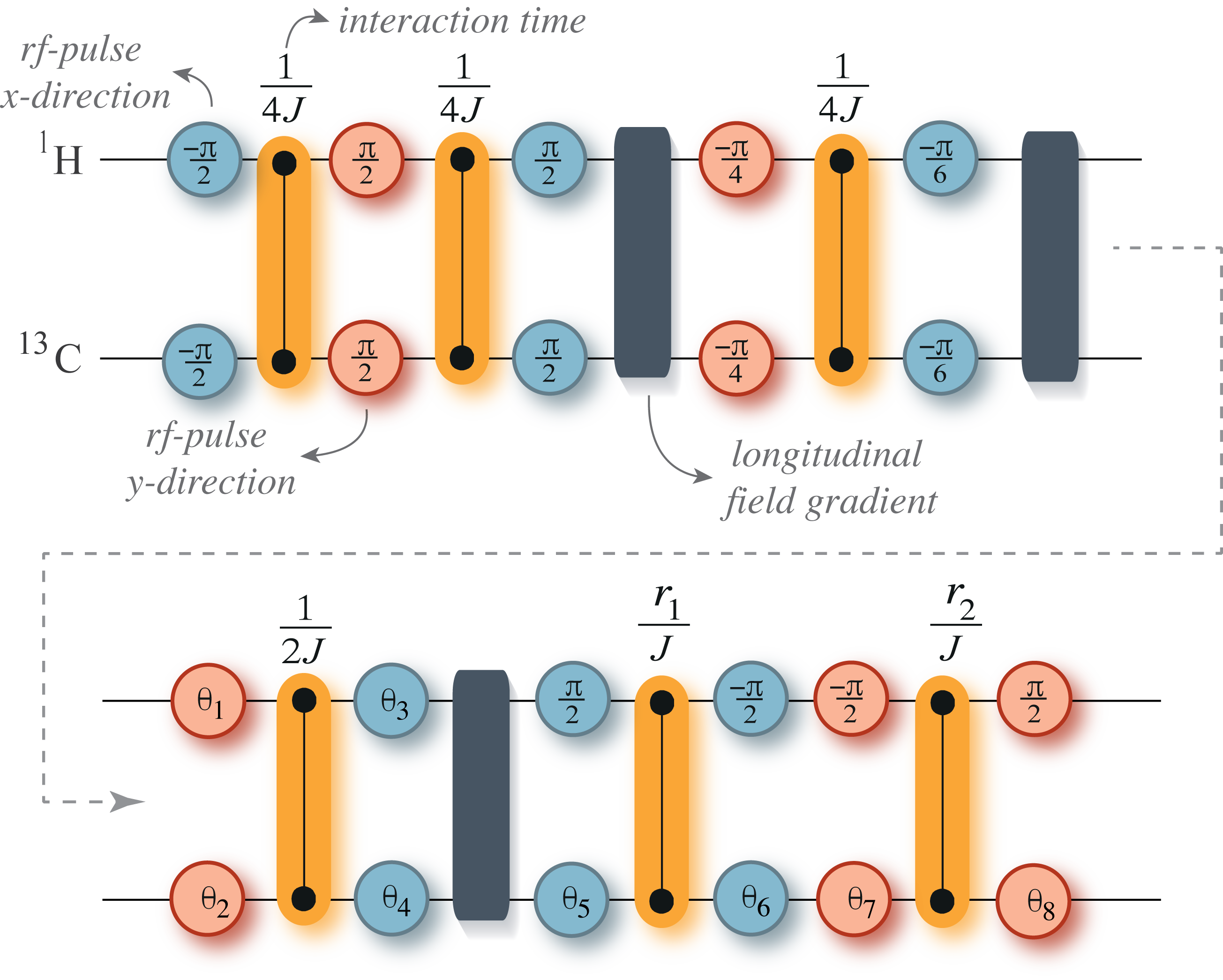} 
{ \caption{\textbf{Pulse sequence for the initial state preparation. }The blue
(red) circle represents $x$ ($y$) local rotations by the indicated
angle. Such rotations are produced by an transverse rf-field resonant
with $^{1}\text{H}$ or the $^{13}\text{C}$ nuclei, with phase, amplitude,
and time duration suitably adjusted. The orange connections represents
a free evolution under the scalar coupling, $\mathcal{H}_{J}^{\text{HC}}=({\pi\hbar}/{2})J\sigma_{z}^{\text{H}}\sigma_{z}^{\text{C}}$
($J=215.1$~Hz), between the $^{1}$H and $^{13}$C nuclear spins
along the time indicted above the symbol. The time modulation and
intensity of the gradient pulse, the angles $\left\{ \theta_{1},\ldots,\theta_{8}\right\} $,
and the parametrized interaction times, $r_{1}$ and $r_{2}$, are
optimized to build an initial state equivalent to the one described
in Equation~(1) of the main text.}}
\label{figS1} 
\end{figure}

\begin{table*}[!ht]
\caption{\textbf{Population and local effective spin temperature of the initial states. }The initial population of the nuclear
spin exited state is displayed in the energy eigenbasis, $p_{\text{A(B)}}(1)=\mathrm{Tr}_{\text{B(A)}}\left(\rho_{\text{AB}}^{0}|1\rangle\!\langle1|\right)$.
It is important to note again that the reduced initial state of the
Hydrogen and Carbon nuclei, $\rho_{i}^{0}$, are diagonal in the energy
basis of $\mathcal{H}_{0}^{\text{H}}$ and $\mathcal{H}_{0}^{\text{C}}$,
irrespective of the presence or not of the initial correlation term
$\chi_{\text{AB}}$. }

\centering

\begin{tabular*}{1\textwidth}{@{\extracolsep{\fill}}@{\extracolsep{\fill}}@{\extracolsep{\fill}}lcccccc}
\hline 
Initial state  & $p_{\text{A}}(1)$  & $p_{\text{B}}(1)$  & $\Re(\alpha)$  & $\Im(\alpha)$  & $\beta_{\text{A}}^{-1}$ (peV)  & $\beta_{\text{B}}^{-1}$ (peV)\tabularnewline
\hline 
\hline 
Uncorrelated  & $0.29\pm0.01$  & $0.22\pm0.01$  & \hspace{0.8em}$0.00\pm0.01$  & $-0.01\pm0.01$  & $4.70\pm0.13$  & $3.30\pm0.07$ \tabularnewline
Correlated ($\varphi\simeq\pi$) & $0.28\pm0.01$  & $0.24\pm0.01$  & $-0.19\pm0.01$  & \hspace{0.8em}$0.00\pm0.01$  & $4.30\pm0.11$  & $3.70\pm0.09$ \tabularnewline
Correlated ($\varphi\simeq-\pi/2$)  & $0.32\pm0.01$  & $0.21\pm0.01$  & $-0.01\pm0.01$  & $-0.17\pm0.01$  & $5.60\pm0.18$  & $3.10\pm0.07$ \tabularnewline
\hline 
\end{tabular*}\label{tabSI} 
\end{table*}

\vspace{0.2 in}
\textbf{Supplementary Note 2: Error Analysis}\\
\vspace{0.1 in}

The main sources of error in the experiments
are small non-homogeneities of the transverse rf-field, non-idealities
in its time modulation, and non-idealities in the longitudinal field
gradient. In order to estimate the error propagation, we have used
a Monte Carlo method, to sample deviations of the quantum sate tomography
(QST) data with a Gaussian distribution having widths determined by
the variances corresponding to such data. The standard deviation of
the distribution of values for the relevant information quantities
is estimated from this sampling. The variances of the tomographic
data are obtained by preparing the same state one hundred times, taking
the full state tomography and comparing it with the theoretical expectation.
These variances include random and systematic errors in both state
preparation and data acquisition by QST. The error in each element
of the density matrix estimated from this analysis is about 1\%. All
parameters in the experimental implementation, such as pulses intensity
and its time duration, are optimized in order to minimize errors.


\vspace{0.1 in}
\textbf{Supplementary Note 3: Geometric Quantum Discord} \\
\vspace{0.1 in}

In order to quantify the quantumness
of the initial correlation in the joint nuclear spin state, we use
the geometric quantum discord~\cite{dak10,gir12}. The latter provides
a useful way to quantify nonclassicality of composed system in a general
fashion. A general two-qubit state $\rho$ can be written in the Bloch
representation as, 
\begin{equation}
\rho=\frac{1}{4}\Big(\mathbf{1}+\sum_{j=1}^{3}x_{j}\sigma_{j}\otimes\mathbf{1}+\sum_{j=1}^{3}y_{j}\mathbf{1}\otimes\sigma_{j}+\sum_{j,k=1}^{3}V_{jk}\sigma_{j}\otimes\sigma_{k}\Big),
\end{equation}
where $\{\sigma_{j}\}$ are the Pauli matrices. The closed form expression
of the geometrical quantum discord for a general two-qubit state is
given by \cite{dak10,gir12} 
\begin{equation}
D_{g}(\rho)=2(\mathrm{Tr}\,\Lambda-\lambda_{max}),\label{eq:discordsup}
\end{equation}
where $\Lambda=\big(\vec{x}\vec{x}^{T}+VV^{T}\big)/4$ and $\lambda_{max}$
is the largest eigenvalue of $\Lambda$. We have evaluated Supplementary Equation~(\ref{eq:discordsup})
using the experimentally reconstructed qubit density operators. Note
that the criticisms, concerning the geometrical quantum discord, put
forward in Supplementary References.~\cite{Piani,Hu} do not apply to our case, since
our two-qubit system is isolated. There is hence no third party that
could apply a general reversible trace-preserving map on one of the
spins that could alter the value of the quantum geometric discord.

\vspace{0.1 in}
\textbf{Supplementary Note 4: The Interaction in the Partial Thermalization Protocol Performs no Work}\\
\vspace{0.1 in}

Following a similar reasoning used in Supplementary References~\cite{Barra2015,DeChiara2018}, we will show that the interaction employed in the partial thermalization protocol performs no work. Our system can be described by a Hamiltonian of the form, 
\begin{equation}\label{H}
\mathcal{H} = \mathcal{H}_\text{A} + \mathcal{H}_\text{B} + \mathcal{V}_\text{AB},
\end{equation}
where $\mathcal{V}_\text{AB}$ is the effective interaction between the subsystems $\text{A}$ and $\text{B}$. 
Due to the type of interaction we are considering and the fact that the qubits are resonant, it follows that our model satisfies strict energy conservation:
\begin{equation}
[\mathcal{V}_\text{AB}, \mathcal{H}_\text{A} + \mathcal{H}_\text{B}] = 0.
\end{equation}
Therefore the effective unitary ($\mathcal{U}_\tau$) implemented by the pulse sequence displayed in Figure~1C of the main text also satisfies strict energy conservation ($[\mathcal{U}_\tau,\mathcal{H}_\text{A} + \mathcal{H}_\text{B}] = 0$).
This means that the energy which enters system $\text{A}$ is always equal to the energy that leaves $\text{B}$; viz,
\begin{equation}
{\langle \mathcal{H}_\text{A}  \rangle}_{t} - \langle \mathcal{H}_\text{A} \rangle_0 = - \left( \langle \mathcal{H}_\text{B} \rangle_t - \langle \mathcal{H}_\text{B} \rangle_0\right),
\end{equation}
where $\langle \mathcal{H}_i \rangle_0 = \text{Tr}(\mathcal{H}_i \rho_i^0)$  is the  energy expectation value of the individual spin $i$ at the initial time and $\langle \mathcal{H}_{i} \rangle_t = \text{Tr}(\mathcal{H}_{i} \rho_i^t)$ is the expectation value at any time $t\in [0,{1}/{2J}]$ ($i=\text{A,B}$). 

The above discussion combined with the usual energy conservation for total Hamiltonian, $\langle \mathcal{H} \rangle_t = \langle \mathcal{H} \rangle_0$, implies that 
$\langle \mathcal{V}_\text{AB} \rangle_t = \langle \mathcal{V}_\text{AB} \rangle_0$. 
That is, no extra energy gets trapped in the interaction term. 
In particular, due to our choice of initial state [introduced in Equation (1) of the main text], it is also true that $\langle \mathcal{V}_\text{AB} \rangle_0 = 0$. Whence, 
$\langle \mathcal{V}_\text{AB} \rangle_t = \langle \mathcal{V}_\text{AB} \rangle_0 = 0$. 
We now use these ideas to connect with the notion of work. 

Let us look to the global dynamics, which is unitary so that there can be no heat dissipated to the rest of the universe.
Work, in this case, comes about from the fact that the Hamiltonian~(\ref{H}) is, strictly speaking, time dependent in a small transient interval when the effective interaction $\mathcal{V}_\text{AB}$ is turned on at time $t =0$ and also when it is turned off in another small transient interval at the final time $t'$. 
In this case, one should more appropriately write 
\begin{equation}
\mathcal{H} = \mathcal{H}_\text{A} + \mathcal{H}_\text{B} + u(t)\mathcal{V}_\text{AB},
\end{equation}
where $u(t) =\theta(t)-\theta(t-t')$ is modelled as the sum of two (unity) Heaviside functions ($\theta(x)$) such that $u(t)$ is $1$ if $0<t<t'$ and $0$ if $t<0$ or $t>0$. 
The mean work performed in the process of turning on and off the interaction between the two spin systems can be unambiguously defined as 
\begin{equation}
\left\langle W\right\rangle =\int_{-\infty}^{\infty}dt\left\langle \frac{\partial \mathcal{H}}{\partial t}\right\rangle _{t}
\end{equation}
Since $\dot u(t)=\delta(t)-\delta(t-t')$ (where $\delta(x)$ is the Dirac delta function), it follows that 
\begin{equation}
\left\langle W\right\rangle  = \langle \mathcal{V}_\text{AB} \rangle_0 - \langle \mathcal{V}_\text{AB} \rangle_{t'} = 0. 
\end{equation}
Here we have used the fact that global and local (strict) energy conservation implies that $\langle \mathcal{V}_\text{AB} \rangle_0 = \langle \mathcal{V}_\text{AB} \rangle_{t'}$. Hence, no work is performed when the transient time for turning on and off the time-independent interaction $\mathcal{V}_\text{AB}$ is sufficiently small to be modelled as (unity) Heaviside functions, which is precisely the case in our experiment. The same arguments also hold for the local rotations employed in the pulse sequence displayed in Figure 1C of the main text. Moreover, as discussed above, the expectation value of the potential is always zero at any time of the evolution for the initial state presented in Equation (1) of the main text.
We notice that the same reasoning also holds when the interaction is not turned off at the end of the measurement, as in our case $\langle \mathcal{V}_\text{AB} \rangle_0 = 0$. 
Thus, whether or not the interaction is turned on or off at the end does not alter the main conclusion that our unitary evolution involves no work.

\vspace{0.1 in}
\textbf{Supplementary Note 5: General Initial Correlations}
\vspace{0.1 in}

In the main text, we have considered the correlation term, $\chi_{\text{AB}}=\alpha|01\rangle\!\langle10|+\alpha^{*}|10\rangle\!\langle01|$,
in Equation~(1) of the main text, with $\alpha\in\mathbb{R}$, such that
it does not commute with the thermalization Hamiltonian, $\mathcal{H}_{\text{AB}}^{\text{eff}}=({\pi\hbar}/{2})J(\sigma_{x}^{\text{A}}\sigma_{y}^{\text{B}}-\sigma_{y}^{\text{A}}\sigma_{x}^{\text{B}})$,
$[\chi_{\text{AB}},\mathcal{H}_{\text{AB}}^{\text{eff}}]\neq0$. Now, let us consider a
more general choice for the amplitude of the correlation term, $\alpha=|\alpha|e^{i\varphi}$
with the complex phase $\varphi$. In this case we note that $\chi_{\text{AB}}$
does not commute with $\mathcal{H}_{\text{AB}}^{\text{eff}}$ for $\varphi\neq\pm\pi/2$.
In all these cases, reversals of the arrow of time do occur. However,
the commutator vanishes for the particular value $\varphi=\pm\pi/2$.
In this specific instance only the uncorrelated part of the initial
state, $\rho_{\text{A}}^{0}\otimes\rho_{\text{B}}^{0}$, is involved in the energy
transfer induced by the thermalization Hamiltonian. As a result, the
initial correlations are thermodynamically inaccessible and no reversal
appears, as seen in the experimental data shown in Supplementary Figure.~\ref{figS2}.
So, $[\chi_{\text{AB}},\mathcal{H}_{\text{AB}}^{\text{eff}}]\neq0$ is a necessary condition
to observe reversals of the heat current.

\begin{figure*}[t]
{\centering \includegraphics[width=0.8\textwidth]{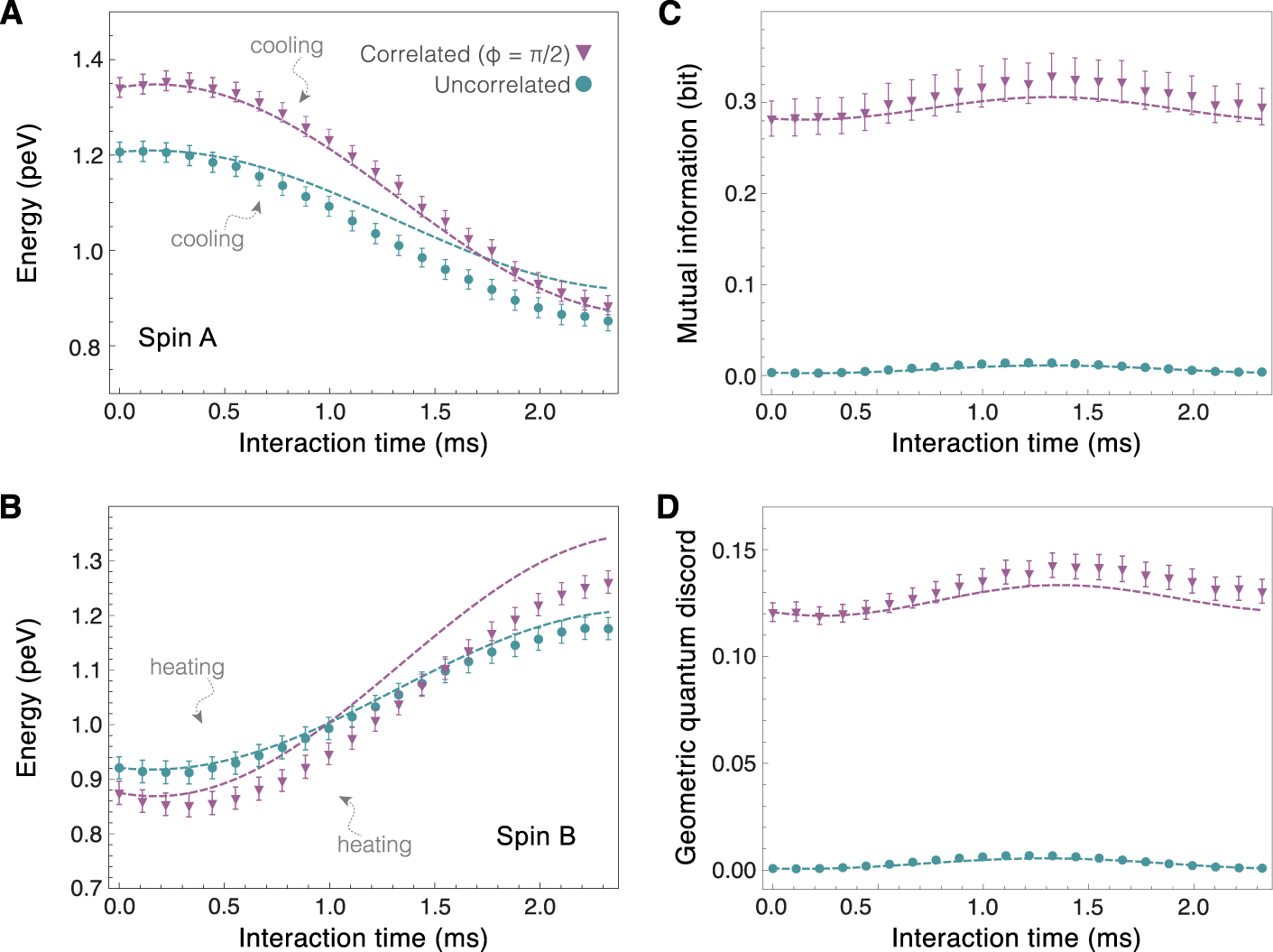}} \caption{\textbf{Dynamics of heat and quantum correlations. }In the absence
of initial correlations, the hot qubit $A$ cools down (\textbf{A})
and the cold qubit $B$ heats up (\textbf{B}). In the presence of
initial quantum correlations that commute with the thermalization
Hamiltonian, $[\chi_{AB},\mathcal{H}_{AB}^{eff}]=0$, the heat current
is not reversed: the initial mutual information (\textbf{C}) and the
geometric quantum discord (\textbf{D}) are not accessible to be consumed
by the thermal interaction. Symbols represent experimental data and
lines are numerical simulations.}
\label{figS2} 
\end{figure*}

\begin{figure*}[t]
\centering \includegraphics[width=0.8\textwidth]{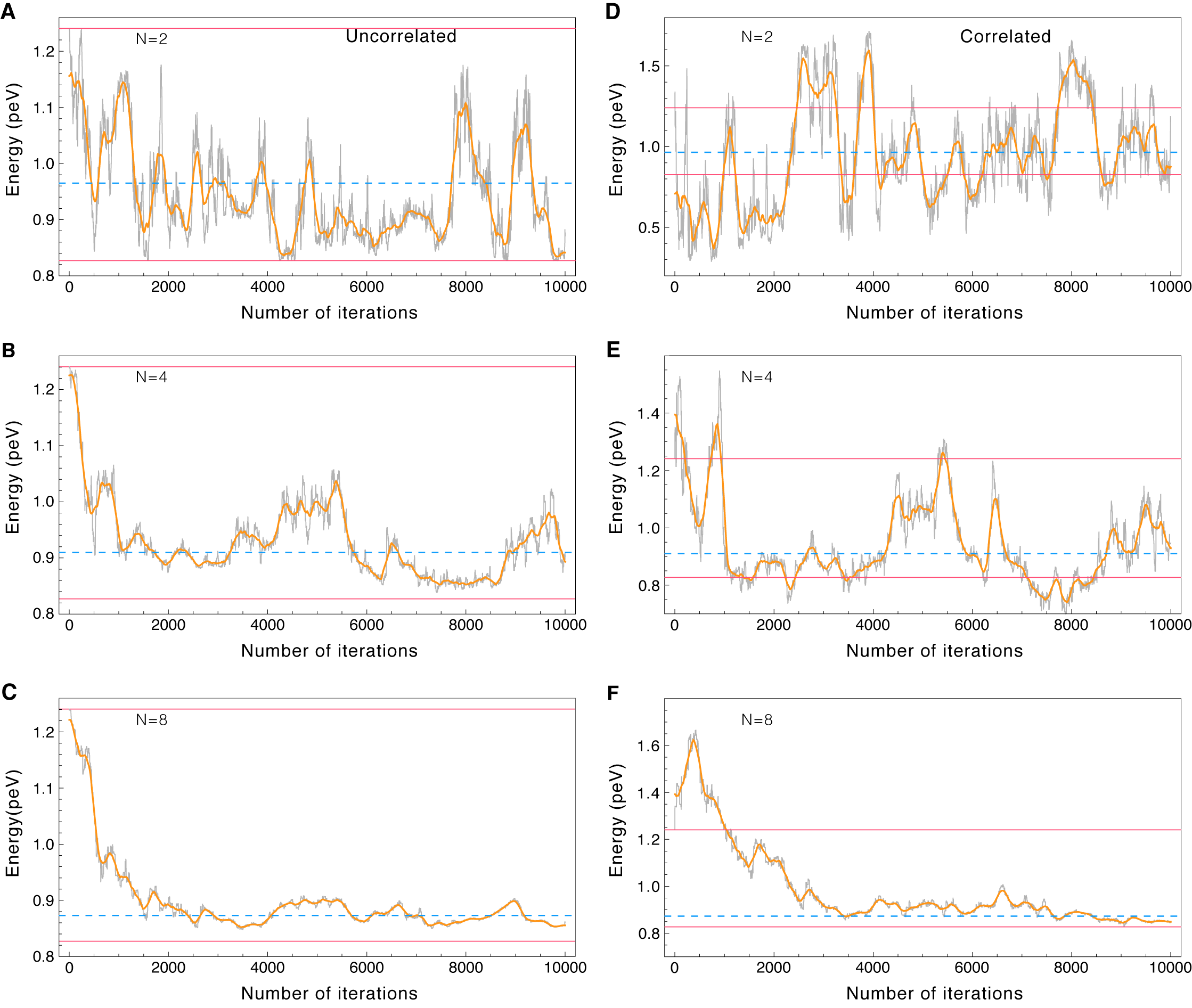} \caption{\textbf{Numerical simulations for larger spin environments. }Average
energy of a system qubit interacting with an ensemble of $N=2,4,8$
bath qubits without initial correlations (\textbf{ABC}) and with initial
correlations (\textbf{DEF}) (grey lines). In both situations, the
system qubit thermalizes to a steady state, corresponding to the average
energy over all the spins, as $N$ increases (blue dashed line). In
the absence of initial quantum correlations, the mean energy of the
system qubit is bounded by the initial mean energies of the hot system
qubit and a cold bath qubit (red solid lines). This corresponds to
the standard arrow of time. However, in the presence of initial quantum
correlations, the mean energy of the system qubit is seen to cross
the red lines. The arrow of time is here reversed as heat flows for
a cold to a hot qubit. These reversals persist even for larger environments
at least for short time dynamics. }
\label{figS3} 
\end{figure*}

\begin{figure*}[t]
\centering \includegraphics[width=0.8\textwidth]{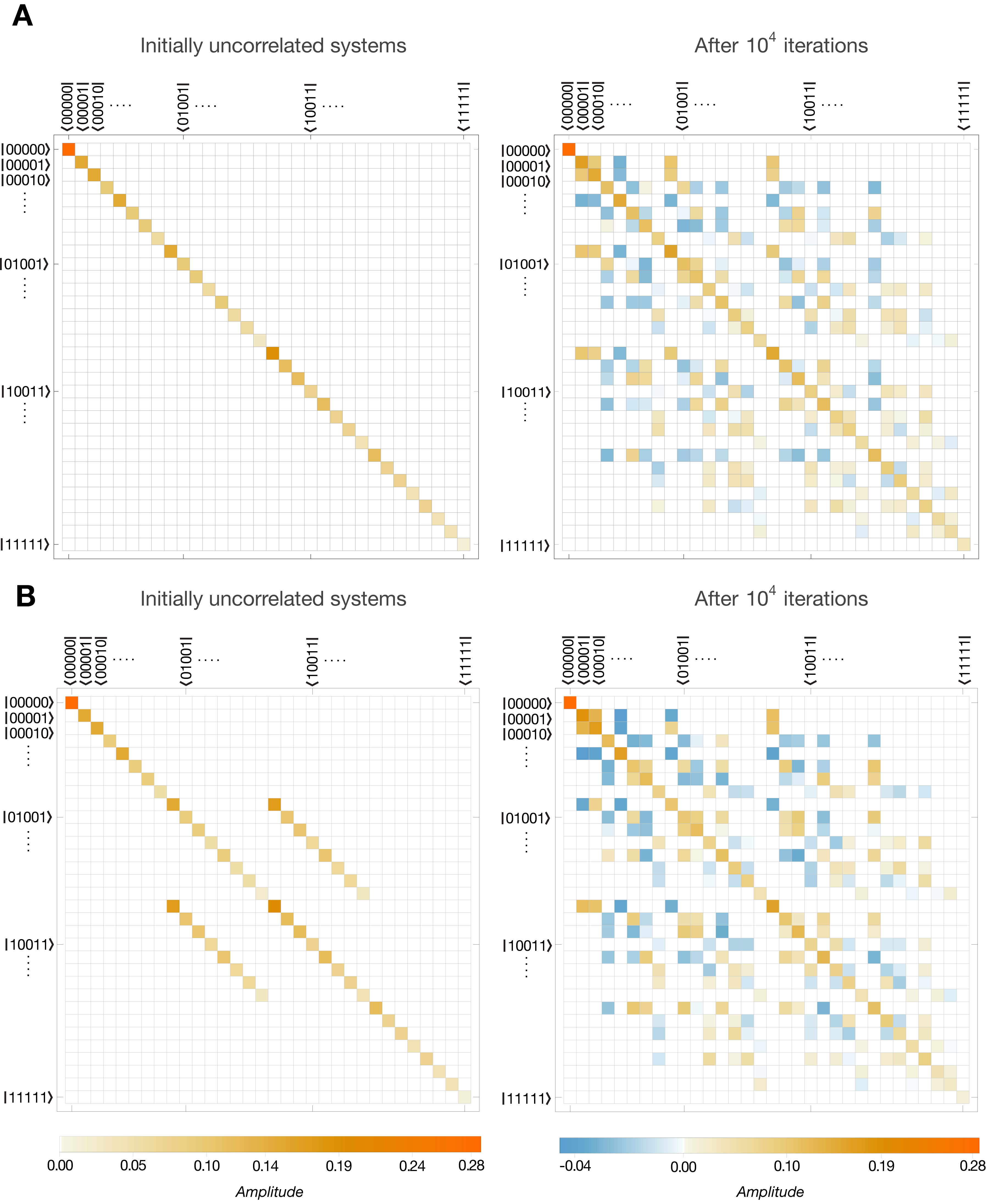} \caption{\textbf{Evolution of the total density matrix elements}. Numerical
simulation of the total density matrix showing thermalization after
$10^{4}$ steps in the initially uncorrelated (\textbf{A}) and correlated
(\textbf{B}) case.}
\label{figS4} 
\end{figure*}

\vspace{0.1 in}
\textbf{Supplementary Note 6: Reversal of the Heat Flux in a Larger Environment} \\
\vspace{0.1 in}

Different thermalization processes for a spin interacting with a multi-spin
environment with random qubit-qubit collisions have been theoretically
investigated \cite{sca02,zim02,ben07}. Supplementary References~\cite{sca02,zim02}
have, for instance, established equilibration induced by individual
collisions with an ensemble of $N$ spins, while Supplementary Reference~\cite{ben07}
has focused on the relaxation generated by repeated collisions with
an ensemble of two spins. Here we will consider, from a theoretical
simulation perspective, a few particle scenario, where each spin,
either from the system or the environment, may interact with any other
spin, much like molecules in a gas. We have concretely considered
a system qubit in an initial state (at hot temperature), $\rho_{0}=\exp(-\beta_{\text{hot}}\mathcal{H}_{0})/\mathcal{Z}_{0}$,
with $(\beta_{\text{hot}})^{-1}=4.881$~(peV), individual nuclear spin Hamiltonian,
$\mathcal{H}_{i}=h\nu(\mathbf{1}-\sigma_{Z}^{(i)})/2$,
and $\nu=1$~kHz as in the main text. The system qubit randomly interacts
with $N$ bath qubits, a bit colder, each one initially in the state
$\rho_{n}=\exp(-\beta_{\text{cold}}\mathcal{H}_{n})/\mathcal{Z}_{n}$, with
$\left(\beta_{\text{cold}}\right)^{-1}=2.983$~(peV), the same individual
nuclear spin Hamiltonian $\mathcal{H}_{n}=\mathcal{H}_{0}$, and $n=1,2,\dots,N$.
The initial state was chosen such that the reduced bipartite density
operator for the qubits $0$ and $1$ reads $\rho_{01}=\mathrm{Tr}_{rest}\,\rho_{\mathrm{total}}=\rho_{0}\otimes\rho_{1}+\alpha(|01\rangle\!\langle10|+|10\rangle\!\langle01|)_{01}$
and all the eigenvalues of the total density operator $\rho_{\text{total}}$
are positive. Here, $\mathrm{Tr}_{\text{rest}}$ denotes the trace over all
the remaining spins except spin 0 and 1. The latter expression is
a direct generalization of the two-qubit case experimentally investigated
in the main text. The random spin-spin collision operator was taken
of the form $U_{\lambda}=\exp[{\lambda(|01\rangle\!\langle10|-|10\rangle\!\langle01|)}]$
\cite{sca02,zim02,ben07} where $|01\rangle\!\langle10|$ act on the
randomly chosen $(j,k)$ spin pair and the interaction parameter satisfies,
$|\lambda|\ll1$.
We have performed extensive numerical simulations using a so-called
gossip (or epidemic) algorithm \cite{dem87} that consists basically
of the following general steps (described here as a pseudo-code):

{\footnotesize{}1:} Define a number $s$ of steps

{\footnotesize{}2:} \textbf{for each} element in $\{1,\dots,s\}$

{\footnotesize{}3:} \hspace{4ex} Choose randomly a pair $(j,k)$
of qubits

{\footnotesize{}4:} \hspace{4ex} Choose randomly a value for $\lambda$
with a Gaussian distribution $\mathcal{N}(0,\pi/50)$

{\footnotesize{}5:} \hspace{4ex} Interact the qubits $j$ and $k$
using $U_{\lambda}$

{\footnotesize{}6:} \textbf{end for}

Such algorithm is used to spread information in a non-structured quantum
network in order to make that all nodes store the same information
\cite{sio16}. The information we are here interested to spreading
is the average individual qubit state $\overline{\rho}_{l}$ with
energy equal to the total energy divided by the number of qubits,
corresponding to the thermalized steady state. After a sufficient
large number of simulation steps, we expect that all individual qubit
states will be close to the average state $\overline{\rho}$.

The results for the number of steps $s=10^{4}$ and system sizes $N=2,4,8$
are shown in Supplementary Figure.~\ref{figS3} for the uncorrelated $(\alpha=0$)
and the correlated $(\alpha=\sqrt{0.0336})$ cases. For each value
of $N$, we have used the same seed for the pseudo-random number generator,
so that each pair of correlated-uncorrelated simulations compares
two systems under the same discrete evolution history. The grey lines
represent the simulated mean energy of the system spin as a function
of the number of simulation steps. Since the simulations are rather
noisy (especially for small $N$), we have added a smoothed orange
line for better visualization of the results. The dashed blue line
corresponds to the total average energy. We observe in both cases
that the mean system spin energy asymptotically relaxes to the total
average energy as $N$ increases, as expected. The red solid lines
(in Supplementary Figure.~\ref{figS3}) indicate the respective average initial energies
of the (hot) system spin $\rho_{0}$ and of the (cold) bath spins
$\rho_{n}$. In the uncorrelated case $(\alpha=0$), the mean system
spin energy is always bounded by the two average initial energies.
Here, heat always flows from the hot to the cold spins on average.
By contrast, for the correlated case $(\alpha=\sqrt{0.0336})$, the
mean system spin energy is seen to cross the red lines (the upper
of lower bound of the standard case), revealing a reversal of the
arrow of time along the evolution steps.

We may understand how the random interactions induce relaxation by
looking to one state of the case $N=2$. We focus on one interaction
$U_{\lambda_{2}}^{(1,2)}$ between spins $1$ and $2$ that takes
place after one previous interaction $U_{\lambda_{1}}^{(0,1)}$ between
spins $0$ and $1$. Since $|\lambda_{1}|,|\lambda_{2}|\ll1$, we
may apply the Baker-Hausdorff formula to obtain, 
\begin{eqnarray}
U_{\lambda_{2}}^{(1,2)}U_{\lambda_{1}}^{(0,1)} & = & \exp(\lambda_{1}\mathcal{H}^{(0,1)}+\lambda_{2}\mathcal{H}^{(1,2)}\nonumber \\
 &  & +\frac{\lambda_{1}\lambda_{2}}{2}\mathcal{Z}^{(1)}\otimes\mathcal{H}^{(0,2)}),
\end{eqnarray}
where $\mathcal{H}^{(i,j)}=\left(|01\rangle\!\langle10|-|10\rangle\!\langle01|\right)_{(i,j)}$.
Since $\rho_{1}$ is not a fully mixed state, the term proportional
to $\mathcal{Z}^{(1)}$ will induce correlations between $\rho_{0}$
and $\rho_{2}$ due to interference effects. However, as $N$ increases,
the probability that the same pair randomly interacts twice in a row
decreases significantly. As a result, a large number of interactions
will create an apparent dephasing in the subspace of each pair, at
the same time as the total global correlations between all the spins
increase, see also Supplementary Figure~\ref{figS4}. 

\def\bibsection{\section*{Supplementary References}}

\end{document}